\documentclass[a4paper,twocolumn,english,prl,superscriptaddress,showpacs,longbibliography]{revtex4-1}
\usepackage[latin9]{inputenc}
\usepackage{textcomp}
\usepackage{amsmath}
\usepackage{amssymb}
\usepackage{graphicx}

\makeatletter

\pdfpageheight\paperheight
\pdfpagewidth\paperwidth

\usepackage[colorlinks,citecolor=darkblue,linkcolor=darkred,urlcolor=darkblue] {hyperref}

\usepackage{babel}
\usepackage{xcolor}

\usepackage[backgroundcolor=green!50!white]{todonotes}

\def\be{\begin{equation}} \def\ee{\end{equation}}

\def\bea{\begin{eqnarray}} \def\eea{\end{eqnarray}}

\definecolor{darkblue}{rgb}{0.1,0.2,0.6} \definecolor{darkred}{rgb}{0.8,0.1,0.2}
\usepackage[all]{hypcap}

 \adddialect\l@English\l@english 

\renewcommand{\BibitemShut}[1]{}

\makeatother

\usepackage{babel}
\begin{document}


\title{Anomalous thermalization in ergodic systems}

\author{David J. Luitz}

\affiliation{Department of Physics and Institute for Condensed Matter Theory,
University of Illinois at Urbana-Champaign, Urbana, Illinois 61801,
USA}
\email{dluitz@illinois.edu}

\author{Yevgeny Bar Lev}

\affiliation{Department of Chemistry, Columbia University, 3000 Broadway, New
York, New York 10027, USA}
\begin{abstract}
It is commonly believed that quantum isolated systems satisfying the
eigenstate thermalization hypothesis (ETH) are diffusive. We show
that this assumption is too restrictive, since there are systems that
are asymptotically in a thermal state, yet exhibit anomalous, subdiffusive
thermalization. We show that such systems satisfy a modified version
of the ETH ansatz and derive a general connection between the scaling
of the variance of the offdiagonal matrix elements of local operators,
written in the eigenbasis of the Hamiltonian, and the dynamical exponent.
We find that for subdiffusively thermalizing systems the variance
scales more slowly with system size than expected for diffusive systems.
We corroborate our findings by numerically studying the distribution
of the coefficients of the eigenfunctions and the offdiagonal matrix
elements of local operators of the random field Heisenberg chain,
which has anomalous transport in its thermal phase. Surprisingly,
this system also has non-Gaussian distributions of the eigenfunctions,
thus directly violating Berry's conjecture.
\end{abstract}
\maketitle
Recently, the long standing question of thermalization in closed quantum
systems \cite{VonNeumann1929,*neumann_proof_2010} has regained importance
due to advances in cold atoms experiments \cite{bloch_many-body_2008},
as well as the theoretical prediction of a dynamical phase transition,
known as the many-body localization (MBL) transition between ergodic
and nonergodic phases \cite{anderson_absence_1958,basko_metalinsulator_2006,nandkishore_many-body_2015,altman_universal_2015,vasseur_nonequilibrium_2016}.
Thermalization in classical systems is normally associated with their
underlying ergodicity, a property which is one of the basic assumptions
of statistical mechanics. The situation for quantum systems is more
delicate, since the evolution of any eigenstate amounts to a time
dependent global phase (see recent reviews \cite{yukalov_equilibration_2011,DAlessio2015,borgonovi_quantum_2016}).
Major progress was achieved by Berry who conjectured \cite{Berry1977}
that the coefficients of high energy eigenstates of a quantum system
in a generic basis corresponding to a chaotic classical system are
independent Gaussian variables, similarly to the distribution of the
eigenstates in the corresponding random matrix ensemble \cite{mehta}.
The connection between random matrix theory and realistic systems
was made in Deutsch's seminal paper \cite{deutsch_quantum_1991},
showing that perturbing the Hamiltonian with a random matrix leads
to thermalization. Later, it was shown by Srednicki for a gas of hard
core particles that if Berry's conjecture is satisfied, the distribution
of the velocities of the particles approaches the Maxwell-Boltzmann
distribution for large systems. It was therefore concluded that the
validity of Berry's conjecture is \emph{required} for thermalization
in quantum systems \cite{Srednicki1994}. Building on this intuition,
and the analogy to random-matrix theory, Srednicki proposed that an
ergodic isolated quantum system should satisfy the Eigenstate Thermalization
Hypothesis (ETH) anzatz \cite{Srednicki1995},
\begin{equation}
\left\langle \alpha\left|\hat{O}\right|\beta\right\rangle =\bar{O}\left(E\right)\delta_{\alpha\beta}+e^{-S\left(E\right)/2}f\left(E,\omega\right)R_{\alpha\beta},\label{eq:eth_anzatz}
\end{equation}
where $\alpha,\beta$ are the eigenstates, $\hat{O}$ is a generic
operator, $S\left(E\right)$ is the microcanonical entropy, $\bar{O}\left(E\right)$,
$f\left(E,\omega\right)$ are smooth functions of their arguments,
and $E=\left(E_{\alpha}+E_{\beta}\right)/2$ and $\omega=E_{\beta}-E_{\alpha}$.
Here, the normal distribution with zero mean and unit variance of
the random term $R_{\alpha\beta}$ is justified through Berry's conjecture.
The first, diagonal term in the ETH ansatz is equal to the micro-canonical
expectation value of the corresponding observable, thus representing
a static thermodynamic quantity. This relation was numerically verified
by Rigol et al. for certain generic quantum systems \cite{rigol_thermalization_2008}.
The exponential decay with system size of the second term, as well
as the validity of the Gaussian distribution of the noise $\left(R_{\alpha\beta}\right)$,
was subsequently verified for a number of generic quantum systems
\cite{rigol_alternatives_2012,steinigeweg_eigenstate_2013,ikeda_finite-size_2013,beugeling_finite-size_2014,beugeling_off-diagonal_2015,alba_eigenstate_2015,mondaini_eigenstate_2016,luitz_long_2016}.
In the present work we show that there is a class of ergodic systems
which exhibit anomalous (non-diffusive) relaxation to equilibrium
while still satisfying a modified ETH ansatz, such that the offdiagonal
elements in (\ref{eq:eth_anzatz}) include a power law correction
to their scaling with the system size. To characterize the approach
to equilibrium, we follow the derivations in Refs. \cite{srednicki_approach_1999,khatami_fluctuation-dissipation_2013}
and \cite{DAlessio2015} (Sec.~6.8), and use the correlator,
\begin{align}
C_{\alpha}\left(t\right) & =\left\langle \alpha\left|\hat{O}\left(t\right)\hat{O}\left(0\right)\right|\alpha\right\rangle =\sum_{\beta\neq\alpha}\left|\left\langle \alpha\left|\hat{O}\right|\beta\right\rangle \right|^{2}e^{i\left(E_{\alpha}-E_{\beta}\right)t}.\label{eq:corr_fun}
\end{align}
where $|\alpha\rangle,\,|\beta\rangle$ are eigenstates and in the
last step we have subtracted the element $\beta=\alpha$ (assuming
a generic system with no degeneracy), to have a correlator with a
vanishing infinite time average. Using (\ref{eq:eth_anzatz}) we have, 

\begin{equation}
C_{\alpha}\left(t\right)=\sum_{\beta\neq\alpha}e^{-S\left(E_{\alpha}+\frac{\omega}{2}\right)}\left|f\left(E_{\alpha}+\frac{\omega}{2},\omega\right)\right|^{2}\left|R_{\alpha\beta}\right|^{2}e^{-i\omega t}.
\end{equation}
For further simplification we replace the sum over eigenstates by
an integral over the density of states, which we write as $e^{S\left(E\right)}$:
\begin{equation}
\sum_{\beta\neq\alpha}\to\int\mathrm{d}E_{\beta}\,e^{S\left(E_{\beta}\right)}=\int\mathrm{d}\omega\,e^{S\left(E_{\alpha}+\omega\right)}.
\end{equation}
The Fourier transform to frequency space yields 
\begin{eqnarray}
C_{\alpha}\left(\omega\right) & = & 2\pi\exp\left[S\left(E_{\alpha}+\omega\right)-S\left(E_{\alpha}+\frac{\omega}{2}\right)\right]\\
 & \times & \left|f\left(E_{\alpha}+\frac{\omega}{2},\omega\right)\right|^{2}\left|R_{E_{\alpha},E_{\alpha}+\omega}\right|^{2}.\nonumber 
\end{eqnarray}
Assuming that $S\left(E\right)$ and $f\left(E,\omega\right)$ are
smooth functions of energy and frequency we can expand,
\begin{equation}
S\left(E_{\alpha}+\omega\right)-S\left(E_{\alpha}+\frac{\omega}{2}\right)=\frac{\partial S}{\partial E}\omega-\frac{\partial S}{\partial E}\frac{\omega}{2}=\frac{\omega}{2T}+\mathcal{O}\left(\omega^{2}\right),
\end{equation}
where we used $\partial S/\partial E=1/T$, where $T$ is the micro-canonical
temperature, and we set the Boltzmann constant to one. Expanding the
other term gives 
\begin{equation}
f\left(E_{\alpha}+\frac{\omega}{2},\omega\right)=f\left(E_{\alpha},\omega\right)+\frac{\omega}{2}\left.\frac{\partial f\left(E,\omega\right)}{\partial E}\right|_{E=E_{\alpha}}+\mathcal{O}\left(\omega^{2}\right).
\end{equation}
Therefore to the leading order in $\omega$ we get,
\begin{equation}
C_{\alpha}\left(\omega\right)=2\pi e^{\omega/\left(2T\right)}\left[\left|f\left(E_{\alpha},\omega\right)\right|^{2}+\frac{\omega}{2}\left.\frac{\partial\left|f\left(E,\omega\right)\right|^{2}}{\partial E}\right|_{E=E_{\alpha}}\right].
\end{equation}
For an Hermitian operator, $\hat{O}$ we have $f\left(E_{\alpha},\omega\right)=f\left(E_{\alpha},-\omega\right),$
yielding,
\begin{equation}
\left|f\left(E_{\alpha},\omega\right)\right|^{2}=\frac{1}{4\pi}\left[e^{-\omega/\left(2T\right)}C_{\alpha}\left(\omega\right)+e^{\omega/\left(2T\right)}C_{\alpha}\left(-\omega\right)\right].
\end{equation}
In the limit of small frequencies $\omega/T\to0$, we have,\textbf{
\begin{equation}
\left|f\left(E_{\alpha},\omega\right)\right|^{2}=\int_{-\infty}^{\infty}\mathrm{d}t\,\left\langle \alpha\left|\left\{ \hat{O}\left(t\right),\hat{O}\left(0\right)\right\} \right|\alpha\right\rangle e^{i\omega t},\label{eq:inf_temp_fw}
\end{equation}
}where $\left\{ .,.\right\} $ is an anti-commutator.\textbf{ }We
now \emph{assume} that $\hat{O}\left(t\right)$ is a conserved quantity
which exhibits anomalous transport,
\begin{equation}
\left\langle \Psi\left|\left\{ \hat{O}\left(t\right),\hat{O}\left(0\right)\right\} \right|\Psi\right\rangle \asymp t^{-\gamma}.\label{eq:gamma}
\end{equation}
For such a decay of the correlation function $\left|f\left(E_{\alpha},\omega\right)\right|^{2}$
is given by,
\begin{equation}
\left|f\left(E_{\alpha},\omega\right)\right|^{2}\propto\int_{-\infty}^{\infty}\mathrm{d}t\,\left|t\right|^{-\gamma}e^{i\omega t}\propto\left|\omega\right|^{-\left(1-\gamma\right)}.
\end{equation}
For a finite system of size $L$, saturation will occur after time
$t_{c}\approx L^{1/\gamma}$, analogous to the Thouless time \cite{edwards_numerical_1972}.
This follows from the relation between the return probability exponent
$\gamma$, and the mean-square displacement exponent, which is valid
for one dimensional systems \cite{Alexander1981}. The power-law dependence
will be therefore cut-off for frequencies, $\omega<t_{c}^{-1}=L^{-1/\gamma}$,
and $\left|f\left(E_{\alpha},\omega\right)\right|^{2}$ will become
structureless \cite{DAlessio2015}, 
\begin{equation}
\left|f\left(E_{\alpha},\omega\right)\right|^{2}\approx t_{c}^{1-\gamma}=L^{\left(1-\gamma\right)/\gamma},\qquad\omega<L^{-1/\gamma}
\end{equation}
Then, the offdiagonal elements should scale with system size as,
\begin{equation}
O_{\alpha\beta}\propto e^{-Ls\left(E\right)/2}L^{\left(1-\gamma\right)/\left(2\gamma\right)}R_{\alpha\beta},\qquad\left|E_{\alpha}-E_{\beta}\right|<L^{-1/\gamma}\label{eq:off_diag}
\end{equation}
where we write the micro-canonical entropy density as $s(E)=S(E)/L$,
to make the dependence on system size explicit. Note that we keep
the general form of the ETH ansatz and assume that the distribution
of the random numbers $R_{\alpha\beta}$ has zero mean and unit variance.
The scaling with system size of the standard deviation of the offdiagonal
matrix elements after the dominant exponential factor has been removed
is therefore given by,
\begin{equation}
\text{std}\left(O_{\alpha\beta}e^{Ls\left(E\right)/2}\right)\asymp L^{\delta},\qquad\left|E_{\alpha}-E_{\beta}\right|<L^{-1/\gamma}\label{eq:delta}
\end{equation}
where, $\delta\equiv\left(1-\gamma\right)/\left(2\gamma\right).$
A special case of this relation was established in Ref.~\cite{DAlessio2015}
for diffusive one-dimensional systems, where $\delta=1/2$ and $\gamma=1/2$.
We note in passing, that the scaling of $\left\langle \left|O_{\alpha\beta}\right|^{2}\right\rangle $
with system size was computed in Ref.~\cite{beugeling_off-diagonal_2015}
for generic clean systems and in Ref.~\cite{serbyn_criterion_2015}
for a generic disordered system. In both works, departure from exponential
dependence on system size is observed when $\omega$ \emph{is taken
to be small}. Our results suggest that the cause of this discrepancy
is the logarithmic correction resulting from (\ref{eq:delta}).

To show that (\ref{eq:off_diag}) holds for systems with anomalous
transport, we numerically study the spin\textendash $\frac{1}{2}$
Heisenberg chain in a random magnetic field,

\begin{equation}
\hat{H}=J\sum_{i}\vec{S}_{i}\cdot\vec{S}_{i+1}+\sum_{i}h_{i}\hat{S}_{i}^{z},\label{eq:disorder_Heisenberg}
\end{equation}
where $J$ is the spin-spin coupling, which we will set to 1, and
$h_{i}\in\left[-W,W\right]$ are random fields drawn from a uniform
distribution. Previous studies \cite{bar_lev_absence_2015,agarwal_anomalous_2015,luitz_extended_2016,gopalakrishnan_griffiths_2016,luitz_long_2016,varma_energy_2015,znidaric_diffusive_2016,Khait2016},
have established that the ergodic phase of this model is characterized
by anomalous transport with a continuously varying dynamical exponent
$\gamma\left(W\right)\lesssim1/2,$ as a function of the disorder
strength $W$. The dynamical exponent vanishes at the many body localization
transition, as the system no longer thermalizes in the MBL phase.
In general, exact numerical studies of high energy many-body eigenstates
are a formidable task and full diagonalization becomes very expensive
for systems of size $L\gtrsim18$. Since we strive to access systems
that are as large as possible, we use the shift-invert technique,
which transforms the spectrum of the Hamiltonian such that the states
of interest are moved to the lowest (highest) energies in the transformed
spectrum and become tractable by Krylov space methods. The most commonly
used spectral transformation for this purpose is $\left(H-\sigma I\right)^{-1}$,
where the explicit inversion of the shifted Hamiltonian can be avoided
and replaced by a repeated solution of a set of linear equations.
We use the massively parallel MUMPS library \cite{MUMPS1,MUMPS2}
for this purpose and are able to obtain exact mid-spectrum eigenstates
for system sizes up to $L=22$. For all system sizes, we calculate
a fixed number $k=50$ of eigenstates and eigenvalues in the middle
of the spectrum. For these energy densities the transition to the
MBL phase occurs at a critical disorder strength of $W_{c}\approx3.7$
\cite{luitz_many-body_2015}. In what follows, we will focus on the
limit of small disorder, $W<W_{c}$, where the system is ergodic and
the diagonal elements of local operators were shown to satisfy ETH,
although with non-Gaussian distributions \cite{luitz_long_2016}.
We will show that the \emph{offdiagonal} elements satisfy our scaling
prediction (\ref{eq:off_diag}). Since the many-body density of states
scales exponentially with the system size, for a fixed number of states
around some energy the assumption on the energy difference, $\omega=E_{\alpha}-E_{\beta}=k\exp\left(-Ls\left(E\right)\right)<L^{-1/\gamma}$,
in (\ref{eq:off_diag}) is always satisfied for sufficiently large
systems. 
\begin{figure}[t]
\begin{centering}
\includegraphics[width=8.5cm]{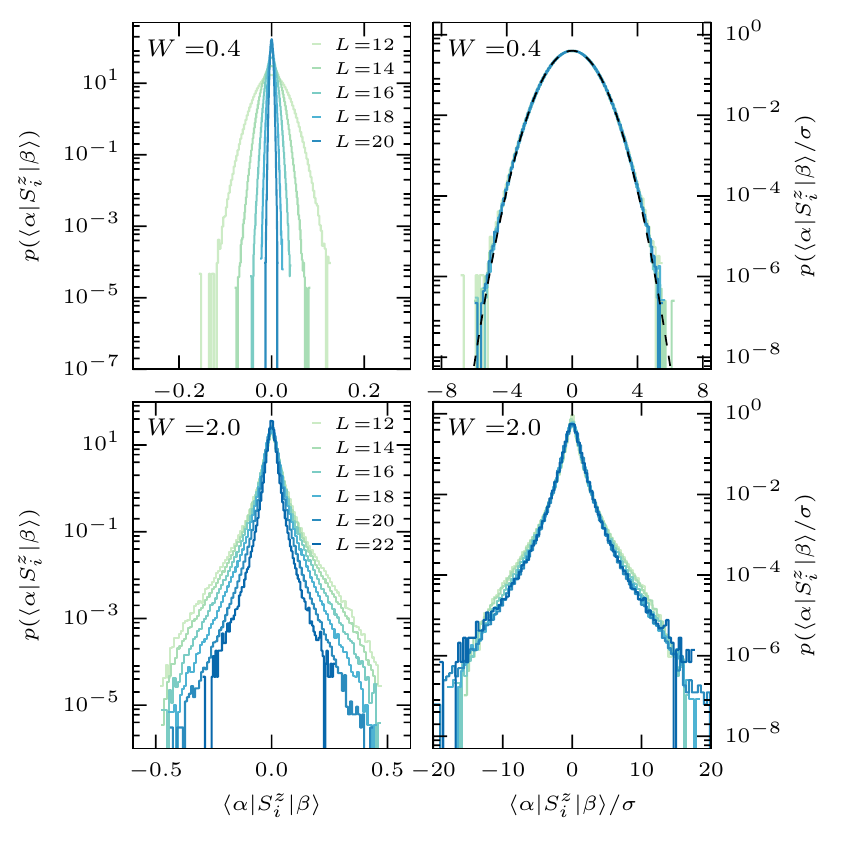}
\par\end{centering}
\caption{\label{fig:off_diag_not_rescaled}\emph{Left column}: Distribution
of the off-diagonal elements ($\alpha\protect\neq\beta$) of the operator
$\hat{S}_{i}^{z}$, written in the eigenstate basis of the Hamiltonian
(\ref{eq:disorder_Heisenberg}), for different disorder strengths,
$W=0.4$ and $2.0$ and system sizes, $L\in[12,22]$. Darker tones
correspond to lager system sizes. The eigenstates correspond to 50
closest eigenvalues to the middle of the many-body spectrum and the
distributions have been sampled from roughly $1000$ disorder realizations,
except for $L=22$, where we only used $100$ realizations. \emph{Right
column}: Distributions rescaled to have a unit variance. At $W=0.4$
the distribution is very close to Gaussian (dashed line).}
\end{figure}

For every pair $|\alpha\rangle,\,|\beta\rangle$ of these eigenstates
with $\alpha\neq\beta$, we calculate the matrix elements $\langle\alpha|\hat{S}_{i}^{z}|\beta\rangle$
of the local $\hat{S}_{i}^{z}$ operator \emph{for all sites $i$
}in the chain using periodic boundary conditions. In the left column
of Fig.~\ref{fig:off_diag_not_rescaled} we present the probability
distribution of the off-diagonal elements computed for different disorder
strengths and system sizes, the right panel shows the same distributions,
renormalized by their standard deviation $\sigma$, in order to compare
the shapes of the distributions across system sizes. This normalization
procedure allows us to directly extract $R_{\alpha\beta}$, since
the resulting distribution has a unit variance. The shape of the rescaled
distribution is Gaussian deep in the ergodic phase (for weak disorder)
and thus corresponds to the general expectation of the ETH ansatz
\cite{steinigeweg_eigenstate_2013,beugeling_off-diagonal_2015}. Closer
to the MBL transition the shape of the distribution is clearly non-Gaussian,
which hints on the violation of the Berry's conjecture. To directly
test the validity of Berry's conjecture we calculate the distribution
of the coefficients $\langle i|\alpha\rangle$ of the eigenfunctions
$|\alpha\rangle$ in the spin basis $|i\rangle$. Surprisingly even
for the smallest disorder we study $\left(W=0.4\right),$ Berry's
conjecture is clearly violated. 
\begin{figure}
\begin{centering}
\includegraphics[width=8.5cm]{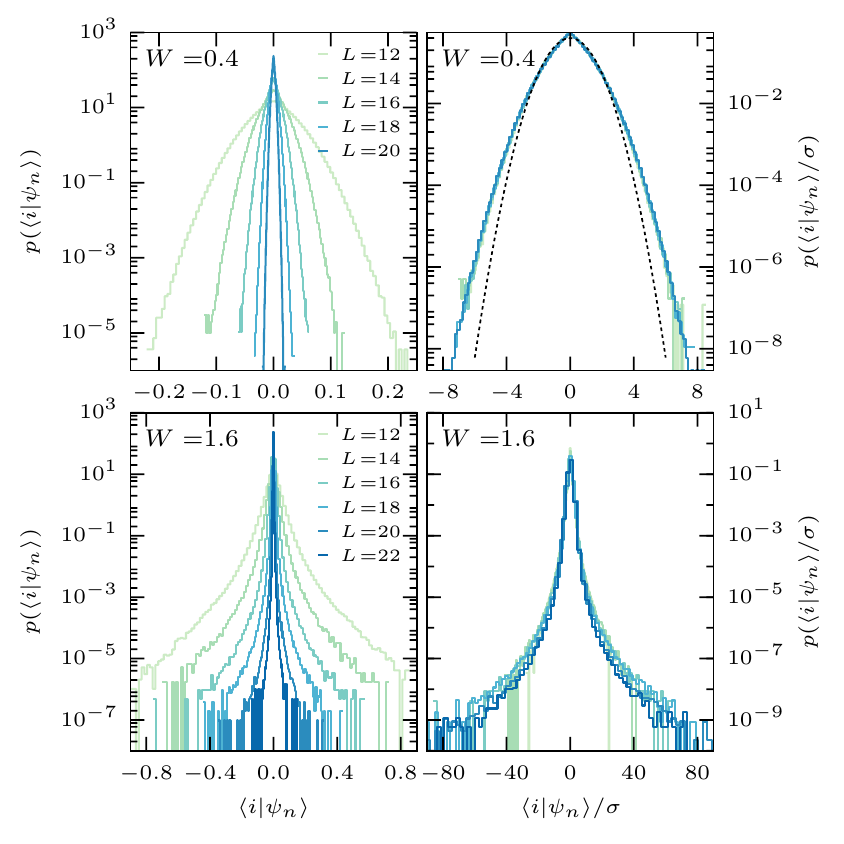}
\par\end{centering}
\caption{\label{fig:berry} \emph{Left column: }Distributions of the eigenfunction
elements in the basis of the local magnetization for small $\left(W=0.4,\text{ top}\right)$
and intermediate ($W=1.6$, bottom) disorder strengths for various
system sizes. \emph{Right column:} Same distribution as in the left
column, rescaled such that the variance is equal to one. Darker tones
correspond to larger system sizes. Clearly, the distribution differs
strongly from a Gaussian distribution at intermediate disorder. At
weak disorder, the difference from Gaussian (dashed line) is visible
mostly in the tails and the excess of weight at zero.}
\end{figure}

\begin{figure}
\begin{centering}
\includegraphics[width=8.5cm]{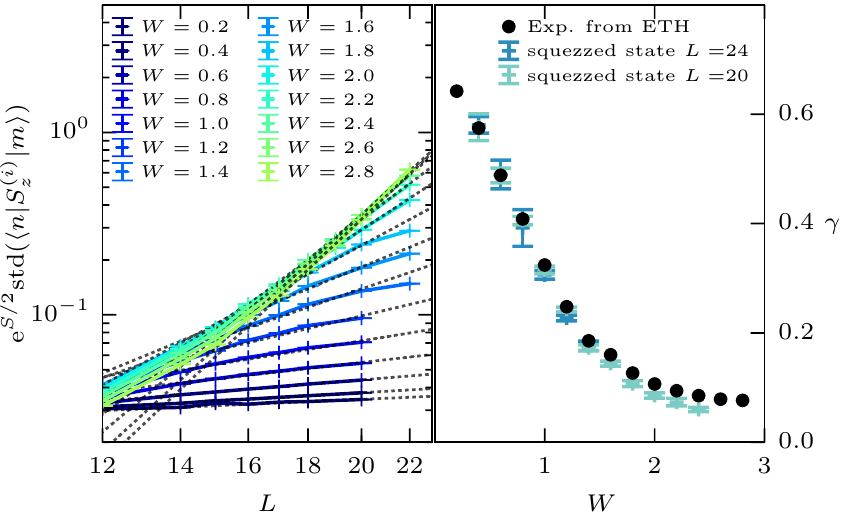}
\par\end{centering}
\caption{\label{fig:exponent} \emph{Left panel}: Extraction of the exponent
from the scaling relation (\ref{eq:off_diag}) for various disorder
strengths after the dominant exponential scaling term was eliminated.
\emph{Right panel}: Exponent extracted from the scaling relation (black
circles) versus direct computation of the dynamical exponent $\gamma$
from the correlation function using energy squeezed states. }
\end{figure}

To verify that the exponent obtained from rescaling according to (\ref{eq:off_diag})
is indeed linked to the dynamical exponent $\gamma$ , we study the
behavior of the correlation function $\langle\psi|\left\{ \hat{S}_{i}^{z}(t),\hat{S}_{i}^{z}\right\} |\psi\rangle$.
As it is very difficult for large systems to obtain high energy eigenstates,
we use random states with an average energy density of $\epsilon=0.5$,
corresponding to the energy $\langle\psi|H|\psi\rangle=E_{\frac{1}{2}}:=\left(E_{\text{max}}+E_{\text{min}}\right)/2$
and a small variance of the energy $\left(\left\langle H^{2}\right\rangle -\left\langle H\right\rangle ^{2}\right)/\left\langle H\right\rangle ^{2}\ll1$.
We generate such typical high energy states starting from a random
state $|\psi_{0}\rangle$ and using the power method for the \emph{folded
}Hamiltonian $(H-E_{\frac{1}{2}})^{2}$ to iteratively reduce the
uncertainty in the energy around $E_{\frac{1}{2}}$. Typically, a
few hundred iterations suffice to reduce the standard deviation of
the energy to a few percent of the bandwidth. We then use the resulting
\emph{energy squeezed states} in the calculation of the correlation
function, which is obtained using exact time evolution by a Krylov
space method \cite{nauts_new_1983,luitz_extended_2016,varma_energy_2015}.
After a short time transient, this function decays as a power law
superposed by oscillations as observed in previous studies for similar
quantities \cite{fabricius_spin_1998,luitz_extended_2016,bar_lev_absence_2015,agarwal_anomalous_2015}.
We find that the most reliable way of extracting the dynamical exponent
$\gamma$ is by using open boundary conditions (OBC) and studying
the correlation function on one of the boundaries. This yields the
same result as the bulk, but the effect of the other boundary is delayed
compared to other setups, which gives access to longer times for which
bulk transport is observed. For smaller system sizes we have verified
that using the eigenstates as the initial condition $|\psi\rangle$
points to similar results. To reliably extract the dynamical exponent
$\gamma$ it is crucial to fit also the transient behavior which includes
decaying oscillations superimposed onto the power law decay. For this
purpose we use the ansatz proposed in Ref. \cite{luitz_extended_2016},
\begin{align}
C(t) & =ae^{-t/\tau}\cos\left(\omega_{1}t+\phi\right),\label{eq:fit}\\
 & +bt^{-\gamma}\left[1+ct^{-\eta}\sin\left(\omega_{2}t+\phi\right)\right]\nonumber 
\end{align}
yielding excellent fits. In Fig.~\ref{fig:exponent} we present the
dynamical exponent $\gamma$ calculated from (\ref{eq:fit}), together
with the exponent $\gamma$, obtained from the exponent $\delta$
(see (\ref{eq:delta})). The left panel of Fig.~\ref{fig:exponent}
shows the LHS of Eq.~(\ref{eq:delta}) as a function of system size
for various disorder strengths on a log-log scale, demonstrating that
it indeed follows a power law. Here, we have estimated the density
of states $e^{S}$ from the energy interval, in which we find $k$
eigenvalues. Note that approaching the MBL transition, visible deviations
from power law behavior appear, signaling the violation of the scaling
(\ref{eq:off_diag}). However, sufficiently far from the MBL transition
the agreement of the two exponents is remarkable. Surprisingly, while
Berry's conjecture \emph{is} violated, the excellent collapse between
the two exponents as predicted by (\ref{eq:off_diag}) suggests that
the ETH anzatz (\ref{eq:eth_anzatz}) still applies, just with \emph{non-Gaussian}
fluctuations and with a modified scaling of the offdiagonal elements
with the system size.

In summary, we have shown that there are systems which are thermal
and exhibit anomalous transport of conserved quantities, but still
satisfy ETH, though in a modified form. We have derived the dependence
of the standard deviation of offdiagonal matrix elements of local
operators (written in the basis of the eigenstates of the Hamiltonian)
on the system size for systems with both normal and \emph{anomalous}
transport. This dependence includes power law corrections to the customary
exponential ETH term. We have derived a scaling relation between the
exponent $\delta$ of this power law, and the dynamical transport
exponent $\gamma$, and thoroughly tested the validity of this scaling
using extensive numerical calculations on the random field Heisenberg
model in its thermal phase. The scaling relation works perfectly for
low to indeterminate disorder strengths sufficiently far from the
MBL transition. Our numerical results also show that the distributions
of the offdiagonal matrix elements are Gaussian at weak disorder,
where the dynamics is roughly diffusive and become strongly non-Gaussian
for stronger disorder, when the system becomes subdiffusive. These
pathological distributions are accompanied by a violation of Berry's
conjecture, as the distributions of the wave function coefficients
deviate strongly from Gaussian distributions. It would be interesting
to explore the possible connection between anomalous transport and
the violation of Berry's conjecture in future works. In our analysis
we have relied only on the second moment of the distributions of offdiagonal
matrix elements, thus ignoring additional information encoded in its
shape, which will show up in the relation between their moments. A
number of previous studies discussed the existence of an intermediate
phase with multifractal eigenstates \cite{de_luca_anderson_2014,luca_ergodicity_2013,chen_many-body_2015,luitz_many-body_2015,torres-herrera_dynamics_2015,altshuler_non-ergodic_2016,tikhonov_anderson_2016}
and multifractal offdiagonal matrix elements of local operators \cite{monthus_many-body-localization_2016}.
While the non-Gaussian form of the obtained distributions is consistent
with these studies, we leave the detailed exploration of this connection
to a subsequent work. 

Since the exponential dependence on the system size of the offdiagonal
elements stems from the randomness assumption of the eigenfunctions
coefficients, we speculate that the derived power law corrections
follow from residual correlations between these coefficients induced
by the conservation laws of the underlying system. It would be therefore
interesting to see how the obtained corrections are affected by the
number of conserved quantities in the system, a question which we
leave for future studies.
\begin{acknowledgments}
\emph{Acknowledgments.} \textendash{} We thank Achileas Lazarides
for inspiring discussions. YBL would like to thank David R. Reichman
for valuable discussions and pointing out the connection to the Berry's
conjecture. DJL thanks Eduardo Fradkin, Anatoli Polkovnikov and Marcos
Rigol for useful comments and Fabien Alet, Bryan Clark, Nicolas Laflorencie
and Xiongjie Yu for related collaborations. 

This work was supported in part by the Gordon and Betty Moore Foundation\textquoteright s
EPiQS Initiative through Grant No. GBMF4305 at the University of Illinois
and the French ANR program ANR-11-IS04-005-01. The code is based on
the PETSc \cite{petsc-efficient,petsc-user-ref,petsc-web-page}, SLEPc\cite{hernandez_slepc:_2005}
and MUMPS\cite{MUMPS1,MUMPS2} libraries and calculations were partly
performed using HPC resources from CALMIP (grant 2015-P0677) as well
as on Blue Waters. This research is part of the Blue Waters sustained-petascale
computing project, which is supported by the National Science Foundation
(awards OCI-0725070 and ACI-1238993) and the state of Illinois. Blue
Waters is a joint effort of the University of Illinois at Urbana-Champaign
and its National Center for Supercomputing Applications.
\end{acknowledgments}

\bibliographystyle{apsrev4-1}
\bibliography{eth_offdiag,eth_offdiag_bib_david,eth_offdiag_bib_manual}

\end{document}